\documentclass[aps,amsmath,amssymb,twocolumn,prb,longbibliography]{revtex4-2}
\usepackage{float}
\usepackage{xcolor}
\usepackage{graphicx}
\usepackage{array}
\newcolumntype{P}[1]{>{\centering\arraybackslash}p{#1}}

\makeatletter
\def\hlinewd#1{%
	\noalign{\ifnum0=`}\fi\hrule \@height #1 \futurelet
	\reserved@a\@xhline}
\makeatother

\usepackage[unicode=true,colorlinks=true,urlcolor=blue,citecolor=blue]{hyperref}

\begin{document}
	
	\title{Exchange-enhanced spin-orbit splitting and its density dependence for
		electrons in monolayer transition metal dichalcogenides}
	
	\author{Igor~Rozhansky}
	\affiliation{National Graphene Institute, University of Manchester, Manchester M13 9PL, United Kingdom} 
	\author{Vladimir Fal'ko}
	\affiliation{National Graphene Institute, University of Manchester, Manchester M13 9PL, United Kingdom} 
	
	\email{igor.rozhanskiy@manchester.ac.uk}

	\begin{abstract} 
		
		We show that spin-orbit splitting (SOS)
		in monolayers of semiconducting transition metal dichalcogenides (TMDs) is substantially enhanced by electron-electron interaction. This effect, similar to the exchange-enhancement of the electron g-factor, is most pronounced for conduction band electrons (in particular, in MoS$_2$), and it has a non-monotonic dependence on the carrier sheet density, $n$. That is, the SOS enhancement is peaked at the onset of filling of the higher-energy spin-split band by electrons, $n_*$, which also separates the regimes of slow (at $n<n_*$) and fast (for  $n>n_*$) spin and valley relaxation of charge carriers. Moreover, this density itself is determined by the enhanced SOS value, making the account of exchange renormalisation important for the analysis of spintronic performance of field-effect transistors based on two-dimensional TMDs.
	\end{abstract}
	\date{\today}
	\maketitle

	Among all two-dimensional (2D) materials, a special place is taken by  monolayers of transition metal dichalcogenides semiconductors~\cite{Manzeli2017,Regan2022,RevModPhys.90.021001,Ahn2020,OBrien2023,PERKINS2024205,Schaibley2016,doi:10.1021/acsnano.5b05556,rosati2024dimensionalsemiconductorsopticalelectronic} (TMDs, $MX_2$, where $M$=Mo,~W and $X$=S,~Se,~Te). These compounds have conduction and valence band edges at the K-points (corner of hexagonal Brillouin zone) with a strong light-matter interaction, ideal for optoelectronics  applications~\cite{Mak2016,Mueller2018,Ciarrocchi2022,Ye2014}. They also feature a strong spin-orbit interaction, inherited from the constituent transition-metal atoms. When combined with middle-plane mirror-symmetry and inversion asymmetry, this results in a spin-valley locking~\cite{PhysRevLett.108.196802,Xu2014}, which can potentially  lead to long spin- and valley-memories of the resulting Ising-type charge carriers~\cite{Tang2019,PhysRevLett.119.137401,PhysRevMaterials.5.044001,doi:10.1126/sciadv.1700518,PhysRevLett.129.027402}. 
	This determines the interest to monolayer TMDs, implemented in field-effect transistors~\cite{Radisavljevic2011,Liu2019Crested,Daus2021,Kshirsagar2016Dynamic,Sebastian2021,Yuan2013,PhysRevLett.127.047202,10.1063/5.0039766},  
	as a promising platform for spintronic devices~\cite{PERKINS2024205,Ahn2020,https://doi.org/10.1002/andp.201400137,SHI202310006,SONG2023100070}. 
	
	In this respect, the magnitude of spin-orbit splitting (SOS) 
	in the TMD monolayer bands, $\Delta$,  
	is important for the spin- and valley-memory of charge carriers: it determines the carrier density range where only one spin-orbit split band is populated in each valley.
	In this regime  spin-valley locking suppresses spin relaxation~\cite{PhysRevB.90.235429} because the latter  requires simultaneous spin-flip and inter-valley scattering to happen in one quantum process. 
	Due to the orbital composition of the $K$-point valence band states~\cite{Kormányos_2015}, SOS is very large for holes~\cite{Yuan2013,Zhang2015,PhysRevB.91.235202,Katoch2018}, which determines their long spin-valley memories. It is much less for conduction band electrons, so that    
	at high carrier densities the electron Fermi level can reach   
	the upper spin-orbit-split bands, thus, opening independent channels for both intra-valley spin-flips (due to  spin-orbit scattering by  wrinkles~\cite{PhysRevB.88.195417,https://doi.org/10.1002/advs.202200816}) and non-spin-flip inter-valley scattering (e.g., by vacancies or $K$-point phonons), leading to a faster spin and valley relaxation.
	By now, the magnitude of SOS for electrons in monolayer TMDs has been studied  theoretically  in numerous  {\it ab initio} density-functional theory (DFT) simulations~\cite{Kormányos_2015,PhysRevB.88.245436,PhysRevB.93.121107,PhysRevB.107.165101}, reporting a relatively wide spread of SOS values for each $MX_2$ compound, as compared to only  handful of experimental results based on the interpretation of magneto-spectroscopy data~\cite{Robert2020,doi:10.1021/acs.nanolett.6b00748,Stier2016,PhysRevLett.131.116901,PhysRevB.107.245407,PhysRevLett.113.076802,PhysRevLett.113.026803,Lu_2020,doi:10.1021/acs.jpclett.3c02431} and magnetotransport data~\cite{PhysRevLett.121.247701,Gustafsson2018,10.1063/5.0039766,PhysRevB.97.201407}.
	Reported values for both conduction and valence band SOS are summarized in Supplemental Materials~\cite{supp} (see also references ~\cite{Mak2012,Klots2014,PhysRevB.90.195434,Kozawa2014,doi:10.1021/acs.jpclett.3c02431,Ross2013,Ye2014,PhysRevLett.113.076802,doi:10.1021/nn305275h,Katoch2018,Ulstrup2016,Riley2014,PhysRevLett.113.026803,Lu_2020,
		Wu2022,Marinov2017,doi:10.1021/acs.jpclett.3c02431,PhysRevLett.121.247701,PhysRevB.107.245407,PhysRevB.107.245407,Kapuściński2021,Wang2017,Zhang2014} therein).
	\begin{figure}
		\centering
		\hspace{2em}	\includegraphics[width=0.3\textwidth]{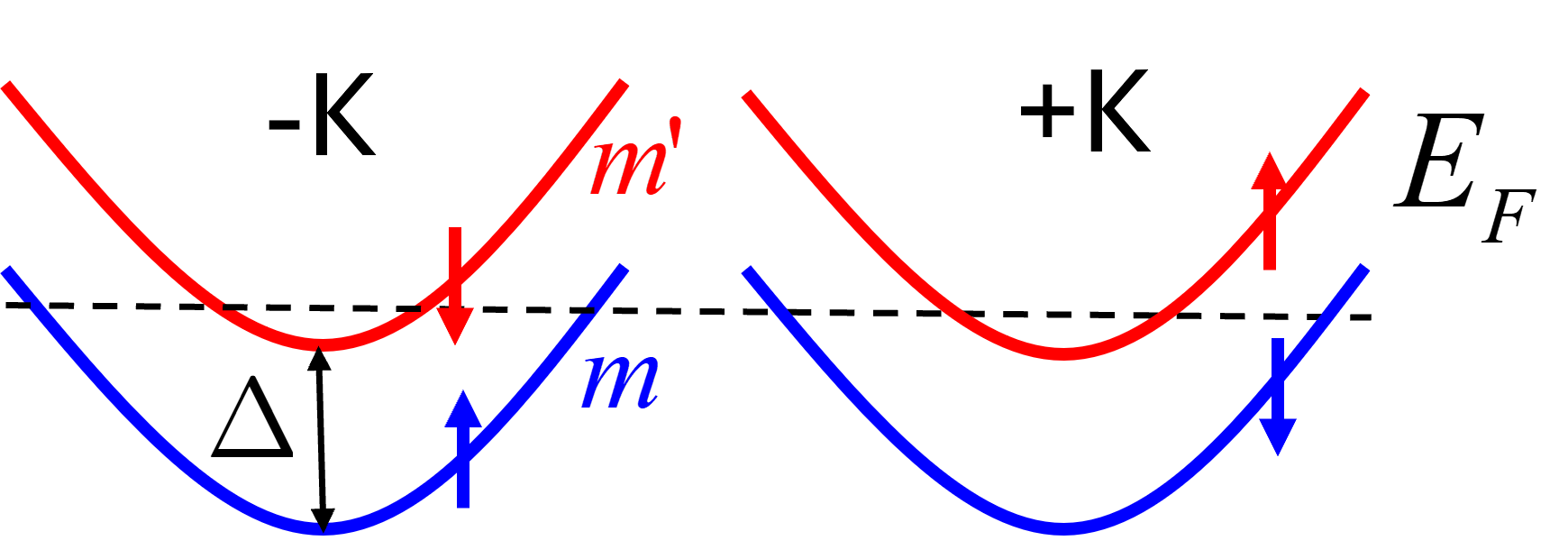}\\
		\includegraphics[width=0.44\textwidth]{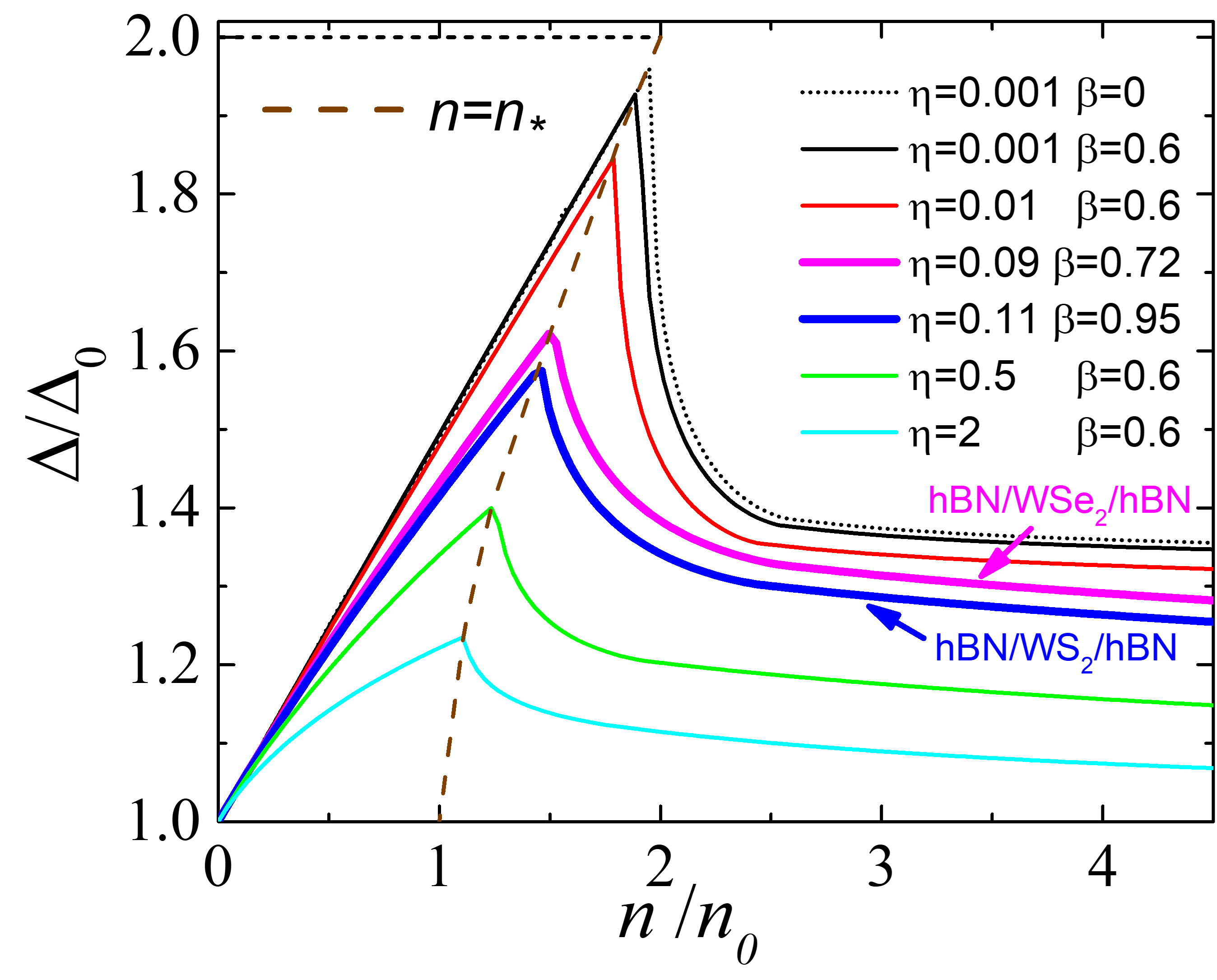}%
		\hspace{1em}
		\vspace{-3ex}
		\caption{{\bf Density-dependent renormalization of spin-orbit splitting, $\Delta$} between two Kramers doublets in TMD monolayers. The data are shown  for different values of bare $\Delta_0$, in-plane  polarisabilities, effective band masses, and various dielectric environments, encoded in $\eta= \frac{\left|\Delta_0\right| }{{{e^2}/{r_*}{\varepsilon}}}$, ${n_0 } = \frac{m \left|\Delta_0\right|}{\pi\hbar^2}$, $\beta  = \frac{{{\hbar ^2}{\varepsilon}}}{{m{e^2}r_*}}\sim  0.6$ (typical for the studied TMDs). The two highlighted curves correspond to WSe$_2$ and WS$_2$ monolayers with parameters determined for the interpretation of exciton spectroscopy data from Ref.~\cite{PhysRevLett.119.047401}.
			Dashed line traces the threshold density,  $n_*$.}
		\label{figsoc}
	\end{figure} 	
	
	Having in mind the above-discussed crossover in the electron spin-valley relaxation in highly doped TMD monolayers, it would be desirable to quantify the relevant threshold density  more precisely. In this respect, we point out that SOS in 2D TMDs, $\Delta(n)$, may be sensitive to the carrier density, $n$, due to exchange interaction between electrons~\cite{Scharf_2019}. This effect is analogous to the exchange enhancement of electron's g-factor established in various 2D systems (such as accumulation layers in semiconductor heterostructures)~\cite{RevModPhys.54.437,PhysRevB.102.125306,CHUDINOV1991961,PhysRevLett.126.067403}. Due to such renormalization, the apparent absolute  value of SOS, $\Delta(n_*)$, at the threshold density, $n_*$, would substantially differ from the single-particle value $\Delta_0$ (provided by DFT computations~\cite{Kormányos_2015,Gjerding_2021,Campi2023}), as illustrated in Fig.~\ref{figsoc} for electrons in the spin-split bands shown in the sketch.
	The dependence $\Delta(n)$ shown in Fig.~\ref{figsoc} is  obtained by analyzing exchange contribution to the self-energy taking into account
	a specific feature of electron-electron interaction in 2D semiconductors~\cite{PhysRevB.84.085406,PhysRevB.91.075310,Goryca2019}: their in-plane dielectric polarisability transforms $1/r$ Coulomb potential into an approximately logarithmic 'Keldysh-like' form~\cite{Keldysh1979CoulombII,rytova2020screened} inside a finite radius, $r_*$, which is longer than a nominal Bohr radius in the system, $r_B$. In Fig.~\ref{figsoc}, we show several exemplary curves for $\Delta(n)$, that we computed for a broad range of input parameters ($r_*>r_B$, the effective electron band mass $m$, $\Delta_0$ defined as always positive absolute value of SOS) and plotted in axis for $\Delta$ normalized by $\Delta_0$ and electron density by 
	${n_{0} } = \frac{m \Delta_0}{\pi\hbar^2}$.  
	Here, the initial rise of $\Delta(n)$ reflects the fact that the exchange energy increases with the number of interacting carriers. This increase is terminated at the density $n_*$, for which the Fermi level of carriers, $E_F$, reaches the higher spin-split bands, $E_F(n_*)=\Delta(n_*)$, and it is followed by a decline, caused by more efficient screening in the systems with a twice-higher density of occupied states and reflecting the form-factor of the Keldysh-like electron-electron interaction.  
	In the set,  displayed in Fig.~\ref{figsoc}, two boldest curves show the results obtained with the input parameters attributed to WSe$_2$ (magenta) and WS$_2$ (blue), for which we obtain  $\Delta_0$ from experimentally measured dark-bright exciton splittings for hBN-encapsulated monolayers~\cite{PhysRevLett.119.047401} following a procedure described in Supplemental Information~\cite{supp} and discussed at the end of this Letter. 
	
	To analyse the renormalization of SOS, we employ the Hartree-Fock theory for the Fermi sea of electrons filling in conduction band of a TMD.
	In the description of the interaction between charge carriers we account for screening by 2D electrons in the random phase approximation (RPA), relevant dielectric environment, as well as the specific feature introduced by the in-plane dielectric polarisability of the TMD crystal. At a single-particle level, electrons near $\tau$K points ($\tau=\pm$) of conduction band edges of a TMD  monolayer are described by
	\begin{equation}
		\label{eqH1} H = \xi_{\bf k}^l + \frac{\Delta_0}{2} \tau s;\quad
		{\xi _{\bf k}^l} \equiv \frac{{{\hbar ^2}{{\bf k}^2}}}{{2m_l}},
	\end{equation}
	where, ${\bf k}\equiv -i{\bf \nabla}_{\bf r}$ is an in-plane wavevector,  and valley and spin states ($z$-component of spin is a good quantum number) are distinguished by  $\tau=\pm 1$ and $\,s=\pm 1$, respectively.
	A bare SOS is described by the energy  difference, $\Delta_0>0$, 
	between two Kramers doublets in Fig.~\ref{figsoc} (upper bands $l=1$ for $\tau s=1$ with $m_1\equiv m'$; lower bands $l=2$ for $\tau s=-1$ with $m_2\equiv m$). For the interaction we use, 
	\begin{align}
		\label{eqvq}
		& V\left( {|\bf{r}}- {\bf{r'}}|\right)=\int {{V_{\bf{q}}}{e^{  i{\bf{q}}\left( {{\bf{r}} - {\bf{r}}'} \right)}}{d^2}{\bf{q}}}/\left( {2\pi } \right)^2 ; \nonumber \\
		& {V_{\bf{q}}} = \frac{{2\pi {e^2}/\varepsilon }}{{q\left( {1 + {r_*}q} \right) + {\Pi _{\bf{q}}}2\pi {e^2}/\varepsilon }};\quad
		{r_*} \equiv 2\pi\kappa/\varepsilon;  \\
		& {\Pi _{\bf{q}}} =  \sum\limits_{\nu=\tau,s}  \frac{m_l}{2\pi {\hbar ^2}}\left[ {1 - \theta \left( {1 - \frac{{8\pi {n_\nu }}}{{{q^2}}}} \right)\sqrt {1 - \frac{{8\pi {n_\nu }}}{{{q^2}}}} } \right],\nonumber
	\end{align}
	as determined by the interplay between dielectric environment with effective dielectric constant $\varepsilon$, polarisability, $\kappa$, of the TMD crystal, and screening by the Fermi sea of the electrons/holes described by polarisation operator $\Pi_{\bf q}$ calculated in the random phase approximation~\cite{Giuliani_Vignale_2005} (which accounts for both spin-orbit split bands of charge carriers -- for details see Supplemental Material~\cite{supp}).  
	
	To describe the renormalization, $\Delta_0$ $\implies$ $\Delta$, in the Hartree-Fock theory, we describe the ground state of electrons as a product of four Slater determinants, build from plane waves, 
	of four spin-valley species.
	Then, we analyse the ground state energy and self-energy contributions to the spectra, 
	\begin{align}
		\label{eqselfen}
		&{\cal E} = \sum\limits_{\tau ,s} {\int {\frac{{d^2{\bf{k}}}}{{{{\left( {2\pi } \right)}^2}}}} {f_{{\bf{k}}\tau s}}\left[ {{\xi _k^l} + \frac{{{\Delta_0}}}{2}\tau s  + \frac{1}{2}{\Sigma _{\tau s}}} \right]} 
		;\nonumber\\
		&{\Sigma _{\tau s}}\left( {\bf{k}} \right) =  - \int {\frac{{d^2{\bf{q}}}}{{{{\left( {2\pi } \right)}^2}}}{f_{({\bf{k}} + {\bf{q}})\tau s}}V_{\bf q}}, 
	\end{align}	
	where $f_{\mathbf{k}\tau s}$ are 
	the Fermi-step functions, corresponding to the densities $n_\nu= \int \frac{{d^2{\bf{k}}}}{{{{\left( {2\pi } \right)}^2}}}{f_{{\bf{k}} \tau s}}$ for each species of particles,  $\nu=(\tau,s)$. 
	
	Then, we minimise ${\cal E}$ with respect to the densities $n_\nu$, under a constraint that $\sum_\nu n_{\nu}=n$, determining the densities, $n_l$, of electrons in each Kramers doublet,  $n_1=n_{+,+}+n_{-,-}$ and $n_2=n_{+,-}+n_{-,+}$ ($n_2$ denotes charge density of the lowest-energy doublet, corresponding to the topmost valence bands or bottommost conduction bands in $K$ and $-K$), which we further use for calculating the SOS renormalization,
	\begin{align}
		\label{eqrenorm}
		&\Delta(n)  = {\Delta _0} + \sum_{l=1,2} (-1)^l \int {\frac{{{V_{\bf{q}}}{d^2}{\bf{q}}}}{{{\left(2\pi\right)}^2}}}  {  \theta \left( {\frac{{{\pi\hbar^2n_{l} }}}{{m_l }} - {\xi^l _{\bf{q}}}} \right)}.
	\end{align}
	We perform these calculations for each of TMDs across a broad carrier density range and combine all relevant material-specific parameters into 
	the following factors,
	\begin{align}
		\label{eqPar}
		n_{0} = \frac{m \Delta_0}{\pi\hbar^2};\,\,
		\eta  = \frac{\Delta_0 }{{{e^2}/{r_*}{\varepsilon}}};\,\,
		\beta  = \frac{{{\hbar ^2}{\varepsilon}}}{{m{e^2}r_*}}\equiv\frac{r_B}{r_*}.
	\end{align}
	The first of these three factors, $n_{0}$, determines a characteristic scale for the crossover density, $n_*$, between the regimes where only the lower  Kramers doublet (with mass $m_2\equiv m$) is filled ($n_2=n<n_*$ and $n_1=0$) and $n>n_*$, where both $n_1$ and $n_2$ are finite. The second, $\eta$, compares bare SOS to Coulomb interaction at the Keldysh length, $r_*$. The third parameter is the ratio between Bohr radius, $r_B=\frac{{{\hbar ^2}{\varepsilon}}}{{m{e^2}}}$, and $r_*$ (in all studied TMD monolayers, $r_B<r_*$). 
	
	In Figure~\ref{figsoc} we use these parameters to represent the results of numerically computed SOS renormalization. At densities $n < n_{0}$, where all electrons are in the lower spin-split band and  screening makes e-e interaction essentially short-range, hence, non-dispersive (in Eq.~\ref{eqvq}, $V_{\bf q}\approx const$ if $\Pi_{\bf q}2\pi e^2/\varepsilon>q, r_*q^2$ for $q\sim k_F$), SOS increases with the density almost linearly, as more electrons are engaged in exchange interaction. This agrees with the earlier discussed long-wavelength limit~\cite{Scharf_2019} of exchange renormalization of the band gap in 2D systems. We note that $\Delta/\Delta_0\le2$ remains always finite, indicating that the system does not undergo any Stoner transition into an Ising ferromagnet. At densities $n \sim n_{0}$, we observe the strongest renormalization, followed by a decline of $\Delta/\Delta_0$, caused by the enhanced screening upon filling the upper spin-split band and the $q$-dispersion of $V_{\bf q}$ at $\Pi_{\bf q}2\pi e^2\varepsilon<q, r_*q^2$.
	
	We also note that for all TMD monolayers $\tfrac{r_B}{r_*} \equiv \beta <1$ (see Table I), hence, we developed a semi-analytical description based on the interaction approximated by,  
	\begin{equation}
		\label{eqvqs}
		{V_{\bf q}} \approx \frac{2\pi e^2/\varepsilon}{{{r_*}{q^2} + \left[1+\theta(n_1)\right]4\pi  e^2/\varepsilon}},
	\end{equation}
	which corresponds to $\beta =0$ and assumes that $m=m'$ as the effect of 10\%-30\% mass difference in the two spin-orbit bands appears to be rather weak  (see Supplemental Information~\cite{supp}). At short distances, the inverse Fourier transform of such $V_{\bf q}$ corresponds to $V(r)\sim (e^2/\varepsilon r_*) \ln(r/r_*)$, characteristic for the Keldysh potential~\cite{rytova2020screened, Keldysh1979CoulombII}, while at long distances it is fully screened. In Fig. 1, we illustrate the validity of such an approximation by comparing the data obtained for $\beta=0.6$ (solid line) and $\beta=0$ (dotted line). This form of $V_{\bf q}$ enabled us to perform the integration in  $k$-space analytically (see Supplemental Material~\cite{supp}), resulting in closed analytical form for the renormalized SOS and densities, 
	\begin{widetext}
		\vspace{-2ex}
		\begin{align}
			\label{eqDelta}
			&	n_1 = \frac{{n - n_{0} }}{2} - \frac{1}{{4{\cal S}}}\ln \frac{{\sqrt {4\left( n - n_1 \right){\cal S}{{\left[ {\theta \left( {{n_1}} \right) + 1} \right]}^{ - 1}} + 1}  + 1}}{{\sqrt {4 n_1 {\cal S}{{\left[ {\theta \left( {{n_1}} \right) + 1} \right]}^{ - 1}} + 1}  + 1}} \text{ for }n>n_*;\quad n_1=0 \text{ for }n<n_* ;\nonumber\\
			& \Delta   \approx  \Delta _0 + \frac{e^2}{2 r_* \varepsilon }\ln \frac{{\left( {n - {n_1}} \right){\cal S} + \left[ {1 + \theta (n_1)} \right]}}{{n_1 {\cal S} + \left[ 1+ \theta (n_1) \right] }};\quad
			{n_*} = {n_{ 0}} + \frac{{\ln \left( {\sqrt {{\cal S}{n_*} + 1/4}  + 1/2} \right)}}{2 {\cal S}}; \quad {\cal S}=\pi r_B r_*.
		\end{align}		
		\vspace{-1.5ex}
	\end{widetext}
	We note that in the limit of metallic  screening, $r_*\rightarrow0$, for which interaction is essentially $\delta$-functional in a real space,   
	$ \Delta  \approx  {{\Delta _0}}  + {E_F}/2$, reproducing the result of  Ref.~\cite{Scharf_2019} obtained for $V_{\bold q}=V_{\bold q->0}$.
	
	We note, that the initial rise of  $\Delta(n)$ in Fig.~\ref{figsoc}
	and in Eq.~\ref{eqDelta} can be used  to describe the SOS renormalization for holes in p-doped TMDs (see Supplemental Information~\cite{supp} for more details), where only densities $n\ll n_0$ can be achieved in realistic devices, due to the much larger values of bare SOS in the valence band.      
	
\begin{figure}
		\centering
		\includegraphics[width=0.4\textwidth]{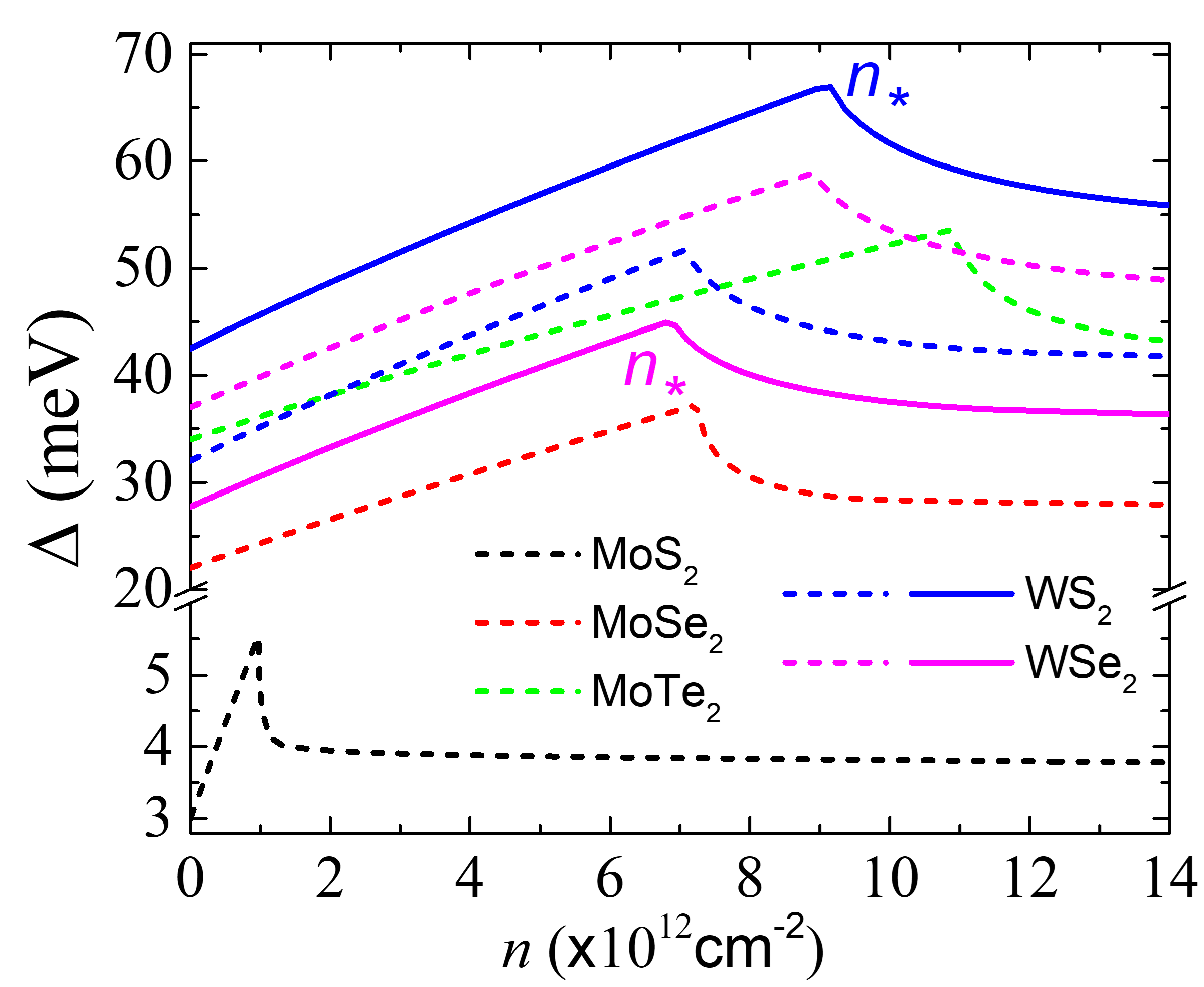}
		\includegraphics[width=0.4\textwidth]{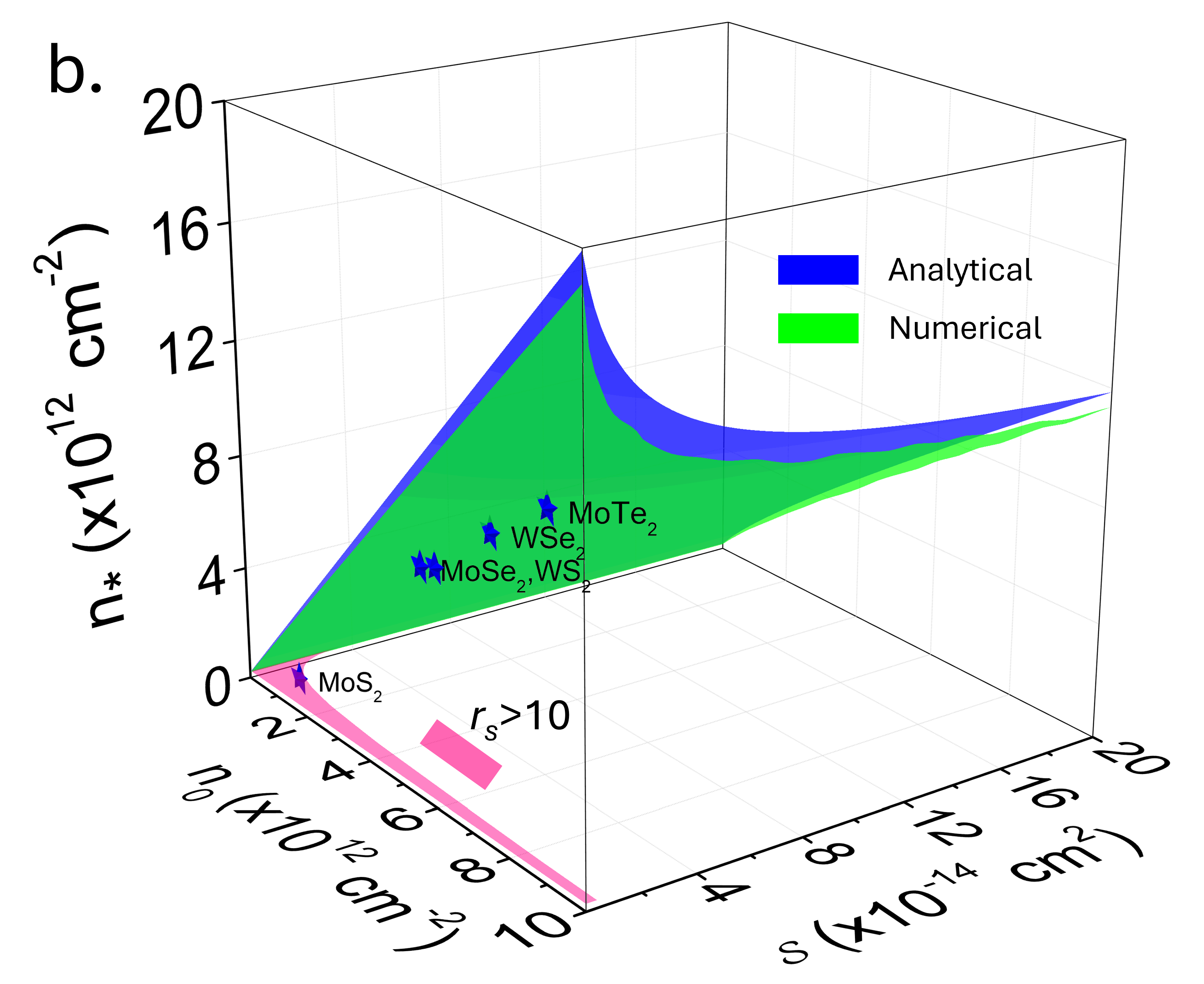}
		\caption{{\bf Exchange-enhanced SOS for electrons in TMD monolayers.} (a) Dashed lines correspond to material parameters  listed in Table~\ref{tabE}. Solid lines show the calculation using single-particle
			SOS values $\Delta_0$ calculated from exciton splittings in W$X_2$ compounds reported in Ref.~\cite{PhysRevLett.119.047401}. The corresponding values of the threshold densities are $n_*=6.8\cdot10^{12} \text{ cm}^{-2}$ for WSe$_2$ and $n_*=9.0\cdot10^{12} \text{ cm}^{-2}$
			for WS$_2$.
			(b) Density $n_*$ calculated analytically using Eq.~(\ref{eqDelta}) (blue surface) and numerically (green), with stars identifying $n_*$ for the studied $MX_2$ compounds. Red region on the bottom plane shows where $r_s>10$.}
		\label{figSOCE}
	\end{figure}

	In Figure ~\ref{figSOCE} we display the results of SOS renormalisation for K-valley electrons for hBN-encapsulated monolayers. 
	We note that the SOS renormalization factor is the largest in MoS$_2$, almost doubling the value of $\Delta$ at the density $n = n_* \approx 10^{12} \text{ cm}^{-2}$. 
	\begin{table}
		\centering
		\begin{tabular}{|P{1.3cm}|P{0.7cm}|P{0.7cm}|P{0.8cm}|P{0.8cm}|P{1.2cm}|P{0.95cm}|P{0.95cm}|}
			\hline
			\rule[0pt]{0pt}{2.2ex} 
			& $m'$\newline($m_e$)& $m$\newline($m_e$)  &  $\Delta_0$ meV
			&  $r_*\cdot\varepsilon$ \newline nm & $n_{0}/10^{12}$ \newline cm$^{-2}$ & $\eta$   & $\,\,\,\beta$\newline  
			\\
			\hline
			
			\rule[0pt]{0pt}{2.2ex} 
			$\text{MoS}_2$	& 
			0.46& 0.43 & 3 & 3.9 & $0.6$  & 0.008  &  0.75  \\
			\hline
			\rule[0pt]{0pt}{2.2ex} 
			$\text{MoSe}_2$	  & 0.56 &0.49 &  22 &   4 & $4.6$ &  0.05 & 0.83  \\ 
			\hline
			\rule[0pt]{0pt}{2.2ex} 
			$\text{MoTe}_2$	& {0.62} & 0.53 & 34  & 7.3 & $7.4$ & 0.17 &  0.33   \\
			\hline
			\rule[0pt]{0pt}{2.2ex} 
			$\text{WS}_2$	& 0.26 & {0.35} & 32  & 3.8 & $4.7$ & 0.08 &  0.95  \\
			\hline
			\rule[0pt]{0pt}{2.2ex} 
			$\text{WSe}_2$ &0.28	& 0.39 & 37  & 4.5 &
			$6.1$& 0.12 &  0.72 \\
			\hline 
			
			\multicolumn{7}{|c|}{
				\rule[0pt]{0pt}{2.2ex} 
				$
				{\varepsilon} = \sqrt {{\varepsilon ^{||}_{\rm hBN}}{\varepsilon^{zz}_{\rm hBN} }}$ 
			} &
			\multicolumn{1}{|c|}{
				\rule[0pt]{0pt}{2.5ex}  
				4.9 } 
			\\ 
			\hline 		
		\end{tabular}
		\caption{{\bf Parameters for TMD manalyers sourced from DFT data}~\cite{PhysRevB.96.075431,Kormányos_2015}.
			Here, $m\,(m')$ is the effective mass in lower (upper) spin-split band, $r_*\cdot\varepsilon\equiv \frac{1}{2}
			({{{{\varepsilon }_{\rm{M}X_2}^{||}} - 1}})c_{{\rm M}X_2},$ 
			where $\varepsilon_{\rm{M}X_2}^{||}$ and $c_{\rm{MX}_2}$ are in-plane dielectric constant and interlayer distance of the corresponding bulk TMD~\cite{PhysRevLett.114.107401}. 	
		}			
		\label{tabE}
	\end{table}	
	Although relative changes of the SOS are smaller in other TMDs, nevertheless, their absolute increase across the density range up to $10^{13}$ cm$^{-2}$ appears to be  substantial, and we note that for all of these materials $\Delta(n)$ features a non-monotonic dependence with a maximum at $n_*$, $n_{0}<n_*<2 n_{0}$ which determines the onset of filling of the upper spin-orbit-split band. 
	In a wide range of material parameters the numerically calculated $n_*$ using the full interaction potential (\ref{eqvq})
	is well approximated by Eq.~(\ref{eqDelta}), as  shown in   Fig.~\ref{figSOCE}b.
	We note, that at small densities one should expect a departure from the RPA approximation used to describe the metallic screening. The pink region shows the range of parameters where the Wigner radius ${r_s} = \frac{{m{e^2}}}{{\varepsilon {\hbar ^2}\sqrt {\pi n} }} >10$ and the RPA calculation would underestimate SOS renormalization and $n_*$. 
	
	For a confident description of SOS dependence on carrier density and the threshold value $n_*$ it would be wise to use experimentally measured  value of bare SOS, $\Delta_0$. In principle, the latter could be measured using the electron spin resonance (which probes zero-momentum collective excitations in the electronic liquid); however, such an experiment is, yet, to be made.  
	Fortunately, $\Delta_0$ can also be extracted from the energy splitting between bright and dark excitons in monolayers, in particular, in W$X_2$ compounds, where the lowest energy exciton is dark. Such experiments have been reported in Ref.~\cite{PhysRevLett.119.047401} making use of unconventional geometries and peculiar selection rules enabling one to observe both dark and bright excitons with splitting $\delta_{BD}=40$ meV for WSe$_2$ and $\delta_{BD}=55$ meV for WS$_2$, both encapsulated in hBN.
	Then, for determining $\Delta_0$ from the  exciton splitting, one has to take into account the difference in the electron masses $m_1,m_2$ in the spin-split bands ($m_2/m_1=1.39$ in WSe$_2$, $m_2/m_1=1.35$ in WS$_2$~\cite{Kormányos_2015,Zhang2017}): the  heavier mass in the lower electron band increases the dark exciton binding energy, thus  making the exciton splitting larger than $\Delta_0$~\footnote{	
		Additionally, we calculate the renormalization of the effective mass at the bands edges of the spin-orbit split Kramers doublets, 
		$$
		\frac{1}{{{\widetilde{m}_l} }} = \frac{1}{m_l} - {\left. {\frac{{{d^2}}}{{d{k^2}}}\int {\frac{{{V_{\bf q}}d^2{\bf{q}}}}{{\left(2\pi\hbar\right)^2}}} \theta \left( {\frac{{{\pi \hbar^2 n_{l} }}}{{m_l }} - {\xi _{{\bf{k}}^l + {\bf{q}}}}} \right)} \right|_{k = 0}};
		$$
		$$\text{   }\,\,\,\,
		\frac{{{\widetilde{m}_l}}}{{{m_l}}} \approx {\left[ {1 + \frac{1}{{2\pi }}\frac{{\theta \left( {{n_l}} \right)}}{{{r_B}q_{n_l} \left( {1 + {r_*}q_{n_l} } \right) + {(2\pi\hbar^2/m_l)}{\Pi _{{q_{n_l}}}}}}} \right]^{ - 1}}, 
		$$
		where 
		${q_{n_l}} = \sqrt {2\pi {n_l}}$
		For the relevant range of parameters for the studied materials, the exchange-driven mass renormalization  appears to be rather small, preserving the qualitative mass difference as found in DFT calculations~\cite{Kormányos_2015}.
	}. Using the description of exciton binding energies in Ref.~\cite{PhysRevB.95.081301} and open source code available in~\cite{PhysRevB.96.075431}, we recalculate the exciton splittings into single-particle SOS~\footnote{Dark excitons were also ignited using strong in-plane magnetic field in WSe$_2$ with the reported slightly higher value $\delta_{BD}=43$ meV. Here we use data from Ref.~\cite{PhysRevLett.119.047401} for our analysis as the exciton splitting could have been affected by the high magnetic field used in Ref,~\cite{Kapuściński2021}} as $\Delta_0=28$ meV for WSe$_2$ and $\Delta_0=43$ meV for WS$_2$. 
	These values, gathered in Table.~\ref{tab2}, are different from those computed using DFT for isolated monolayers, which may be attributed to the influence of hBN encapsulation~\cite{Kapuściński2021}.
	\begin{table}[h]
		\centering
		\begin{tabular}{|P{1.2cm}|P{1.0cm}|P{1.4cm}|P{1.6cm}|P{1.6cm}|P{1.0cm}|}
			\hline
			\rule[0pt]{0pt}{2.2ex} 
			& $\,\,\,\,\Delta_0$ \newline meV&
			$n_{0}/10^{12}$ \newline cm$^{-2}$  &  $n^{th}_{*}/10^{12}$ \newline cm$^{-2}$ &  $n^{exp}_{*}/10^{12}$ \newline cm$^{-2}$  
			& $\,\,\Delta(n_*)$ \newline meV
			\\
			
			\hline
			\rule[0pt]{0pt}{2.2ex} 
			$\text{WSe}_2$	& 28 & 4.0 & 6.8  & 6.6~\cite{Wang2017} & 45 \\
			\hline
			\rule[0pt]{0pt}{2.2ex} 
			$\text{WS}_2$ &43&  6.3 & 9.0  & - &67	 \\
			
			\hline 		
		\end{tabular}
		\caption{{\bf SOS parameters for  hBN-encapsulated WSe$_2$ and WS$_2$ monolayers}.
			Bare SOS was determined from bright-dark exciton splittings in Ref.~\cite{PhysRevLett.119.047401}. The  threshold density, $n_*^{th}$, was computed  using Eq.~(\ref{eqDelta}) and masses given in Table~\ref{tabE}, in comparison with the only available experimental result, $n_*^{exp}$, from Ref.~\cite{Wang2017}. 
		}
		\label{tab2}
	\end{table}
	
	For those  values of $\Delta_0$, the computed $\Delta(n)$ is shown by two  highlighted curves in Fig.~\ref{figSOCE}. 
	In particular, we note that, for WSe$_2$ the enhancement of SOS determines the threshold density $n_*=6.8\cdot10^{12}\text{ cm}^{-2}$ (as compared to $n_0=4.0\cdot10^{12}\text{ cm}^{-2}$ estimated from bare SOS), which is close to the threshold density $n_*=6.6\cdot10^{12} \text{ cm}^{-2}$ observed in Ref.~\cite{Wang2017} based on the analysis of Shubnikov-de Haas oscillations.
	Such a threshold density plays an important role for spintronic characteristics of $MX_2$-based FETs. This is because it separates the regime of slow spin-valley relaxation of electrons at $n<n_*$ (because such relaxation requires simultaneous spin-flip and inter-valley scattering, as prescribed by the spin-valley locking), from the fast spin-relaxation regime at $n>n_*$ (where both intra-valley spin-flips and non-spin-flip inter-valley processes are allowed)~\cite{PhysRevB.90.235429}.  
	
	We thank K. Ensslin, A. Geim, L. Glazman, and M. Potemski for useful discussions. We acknowledge support from EPSRC Grants EP/S019367/1, EP/P026850/1, and EP/N010345/1; British Council project 1185409051; 
	I.R. gratefully acknowledges the support of CARA Fellowship and the University of Manchester.
	
	\bibliography{soc}
	
\end{document}